\documentclass[aps,pra,twocolumn,superscriptaddress,10pt]{revtex4-2}

\usepackage{amsmath, amssymb, amsthm}
\usepackage{mathtools}
\usepackage{mathrsfs}
\usepackage{physics}

\usepackage{graphicx}

\usepackage{algorithm2e}
\RestyleAlgo{ruled}

\usepackage{xcolor}

\usepackage{hyperref}

\hypersetup{
    colorlinks = true,  
    linkcolor = blue,   
    citecolor = blue,   
    urlcolor = blue,    
}

\newcommand{\Hil}{\mathcal{H}}
\newcommand{\Fock}{\mathcal{F}}

\begin{document}

\title{Noise Limits on Fault-Tolerant Fermionic Quantum Computing}

\author{Owen Allison}
\thanks{owen.allison@colorado.edu}
\affiliation{Department of Physics, University of Colorado, Boulder, Colorado 80309, USA}
\affiliation{JILA, NIST, and University of Colorado, Boulder, Colorado 80309, USA}

\author{Luke Coffman}
\affiliation{Department of Physics, University of Colorado, Boulder, Colorado 80309, USA}
\affiliation{JILA, NIST, and University of Colorado, Boulder, Colorado 80309, USA}
\affiliation{Harvard Quantum Initiative, Harvard University, Cambridge, MA 02138, USA}
\affiliation{Department of Physics, Harvard University, Cambridge, MA 02138, USA}
\affiliation{School of Engineering and Applied Sciences, Harvard University, Cambridge, MA 02138, USA}

\begin{abstract}
  Determining the highest amount of noise that quantum circuits can handle is an interesting and crucial task in the development of fault-tolerant quantum computation. Previous work has constrained this upper limit for local depolarizing noise to $\approx 45~\%$ for circuits constructed using the universal Clifford with T gate set by finding the noise threshold where the gate set loses universality. In this work, we use a similar method with the matchgate with non-Gaussian resource universal gate set to find a limit of $p=(8 - 2 \sqrt{6})/5$ or $\approx 62~\%$ for fermionic quantum computing that is independent of circuit depth. To do this, we use Uhlmann-Wootters concurrences for a 4-mode fermionic Choi state representing a resourceful gate combined with local depolarizing noise to determine when the combined channel is convex Gaussian. These bounds are not directly comparable due to differences in noise model making the bound of $\approx 62~\%$ the best known for fermions.
\end{abstract}

\maketitle

\section{Introduction}

The simulation of physical systems, especially chemical systems, has been a key inspiration for the development of quantum computers~\cite{feynman_simulating_1982}. However, an ever-present challenge in this endeavor is noise. While small amounts of noise can preserve genuine quantum effects that cannot be replicated classically, it can still render the system non-universal and/or efficiently computable on a classical computer~\cite{aharonov_limitations_1996}. Although quantum error correction and fault-tolerant constructs can combat this issue, sufficiently high noise can inhibit fault-tolerant encoding altogether~\cite{aharonov_fault-tolerant_2008,devetak_private_2004,bennett_capacities_1997}. Thus, sufficiently high noise can render a system effectively no longer fault-tolerant. Therefore, better understanding the noise limits of different quantum systems remains not only an interesting theoretical question but a crucial task in the development of fault-tolerant quantum computers.

Several works have previously explored the noise limits of different circuit and noise models, a primary goal of which being to find lower bounds below which it is certain that a quantum circuit can be made fault-tolerant through the use of error correcting codes. For example, it was found that fault-tolerant quantum computation is possible when error rates are less than a constant lower bound limit for noise of about $~10^{-6}$ or $~10^{-4}~\%$, a bound later improved to $~0.01~\%$~\cite{aharonov_fault_tolerant_1999, aliferis_accuracy_threshold_2007}. Although useful, lower bounds do not rule out the possibility of fault-tolerant computation at higher levels of noise. Thus, it is of interest to establish an upper bound on noise after which fault-tolerant computation is impossible. Early work showed that circuits consisting of single-qubit and 2-qubit gates with depolarizing noise of $74~\%$ or more were efficiently computable~\cite{harrow_robustness_noise_2003}. Further works found upper bounds of $50~\%$ and $35.7~\%$ for local depolarizing noise in circuits with $\leq$2-qubit gates and uniform noise, and circuits with noisy 2-qubit and effectively noise-free 1-qubit gates respectively after which noise accumulated in circuits of log depth causes the output to become effectively independent of the input and thus effectively useless.~\cite{razborov_upper_bound_2003, kempe_2008_threshold}. 

It is possible for sufficient noise to be applied such that a circuit is rendered non-universal and/or efficiently computable and thus effectively not fault-tolerant regardless of the depth of the circuit. For instance, Clifford circuits are a well known family of efficiently computable operations~\cite{shor_fault_tolerant_1997, gottesman_heisenberg_1998, gottesman_simulation_stabilizer_2004}. After depolarizing noise of $p\approx 0.45$ or $\approx45~\%$, any single qubit state becomes contained within the stabilizer polytope~\cite{buhrman_new_2006}, and thus the resourceful T gate (which is the most resistant to noise) no longer elevates Cliffords to universal quantum computation or non-efficient computability, and so fault-tolerance is impossible. This level of noise therefore serves as an upper bound for circuits constructed with this gate set after which fault-tolerant quantum computation is impossible.

Another class of efficiently computable non-universal circuits is matchgate circuits~\cite{knill_matchgates_2001, valiant_quantum_circuits_2002, terhal_classical_simulation_2002}. Matchgates are of particular interest because they form a continuous family, unlike the Clifford group, and correspond to free-fermionic (fermionic Gaussian) unitary evolution, or unitaries that represent non-interacting fermion evolution. Therefore, they are a natural choice for simulating physical systems for quantum chemistry once they are promoted to universality with the addition of parity-preserving non-Gaussian gates such as SWAP~\cite{Jozsa_matchgates_simulation_2008, brod_matchgates_universal_2011}. Because all of the gates therein are parity preserving, circuits constructed with this gate set correspond to circuits acting on fermionic modes, that is, fermionic quantum computing~\cite{bravyi_fermionic_computation_2002}, and are related via the Jordan-Wigner transform. 

In this paper, we determine an upper bound limit analogous to the $\approx 45~\%$ result for circuits constructed with the matchgate with non-Gaussian resource gate set, and as a result fermionic quantum computation. We use the noise model in which local depolarizing noise is applied after each non-Gaussian gate and each matchgate is assumed effectively noise-free, and also that circuits takes standard non-magic states as input, that is, Gaussian states such as the vacuum. This upper bound is not directly comparable to previous results because it is independent of circuit depth and analyzes a different noise model. For instance, consider the SWAP gate which is equivalent to three consecutive CNOT gates. Implementing a circuit with three CNOT gates taking the place of a SWAP gate and maintaining the same noise model results in each CNOT being followed by depolarizing noise. This is a different noise model than that used in Ref.~\cite{buhrman_new_2006}, and thus the upper bound found therein cannot be directly compared to that found in this paper.

We can determine the upper bound limit by finding the level of noise at which the combined channels consisting of SWAP (or another resourceful gate) and the depolarizing noise become convex Gaussian, that is, they correspond to fermionic channels that map convex combinations of fermionic Gaussian input states to convex combinations of fermionic Gaussian output states, at which point the circuit loses the non-Gaussian support extending it to universality and becomes non-universal. When this happens, the evolution of the circuit can be tracked efficiently by a classical computer~\cite{oszmaniec_classical_2014}. The point at which this first occurs is the point at which fault-tolerant computation becomes effectively impossible. We find this upper bound limit to be $\approx 62~\%$ specifically when SWAP is used as the resourceful non-Gaussian gate. In the appendices, we show this is the maximum for all 2-qubit parity preserving gates with a closed form of exactly $p=(8 - 2 \sqrt{6})/5$.

\section{Background}

In this section, we give background on information pertinent to understanding the contents of the next sections~\cite{nielsen_chuang_2010, altland_simons_2010, surace_scipost_2022}.

\subsection{Fermions}

Fermions are particles with half-integer spins such as electrons, muons, etc. As such, they obey the Pauli exclusion principle, which states that identical fermions may not occupy the same quantum state within a system simultaneously.

\subsubsection{Formalism}

When working with systems of identical fermions, it is often easier to work with occupation numbers for specific single particle states or "modes" rather than spatial or spin states for specific particles. In this notation, a single mode is represented as a state vector $\ket{n}$ where $n$ is the occupation number, or the number of particles in that mode. Because of the Pauli exclusion principle, fermionic modes can only have an occupation number of either 0 or 1. For a system of $k$ modes, a state with a definite number of particles known as a \textit{Fock state} is represented as
\begin{equation}
    \label{eq:fock_state}
    \ket{n_1, n_2, \dots, n_k} = \ket{n_1} \otimes \ket{n_2} \otimes \dots \otimes \ket{n_k}.
\end{equation}

There are $2^k$ possible Fock states that form a basis that spans the Hilbert space of the multi-particle system, the \textit{Fock space} $\Fock_k$. The Fock space can be traversed using the fermionic annihilation and creation operators $a_j$ and $a^\dagger_j$, defined such that
\begin{equation}
    \begin{aligned}
        a_j \ket{n_1, \dots, 1_j, \dots,n_k} &= \ket{n_1, \dots, 0_j, \dots,n_k} \\
        a^\dagger_j \ket{n_1, \dots, 0_j, \dots,n_k} &= \ket{n_1, \dots, 1_j, \dots,n_k}, \\
    \end{aligned}
\end{equation}
and
\begin{equation}
    \begin{aligned}
        a_j \ket{n_1, \dots, 0_j, \dots,n_k} &= 0 \\
        a^\dagger_j \ket{n_1, \dots, 1_j, \dots,n_k} &= 0.
    \end{aligned}
\end{equation}
These operators follow the anti-commutation relation
\begin{equation}
        \{a_i, a_j\} = 0, \quad \{a_i, a^\dagger_j\} = \delta_{i,j}.
\end{equation}

One can alternatively traverse the fermionic Fock space using the \textit{Majorana operators} defined in terms of the annihilation and creation operators as
\begin{equation}
    c_{2j-1} = a_j + a_j^\dagger, \quad c_{2j} = i(a_j - a_j^\dagger),
\end{equation}
which satisfy the anti-commutation relation
\begin{equation}
    \{c_i,c_j\} = 2\delta_{i,j}.
\end{equation}

\subsubsection{Jordan-Wigner Transform}

A single unentangled qubit state can be either $\ket{0}$ or $\ket{1}$. A system of $k$ qubits can then have a computational-basis state
\begin{equation}
    \label{eq:spin_basis}
    \ket{q_1, q_2,\dots,q_k} = \ket{q_1} \otimes \ket{q_2} \otimes \dots \otimes \ket{q_k}.
\end{equation}
It is simple to see then that these are the $2^k$ basis vectors for the qubit Hilbert space $\Hil^k = (\Hil^1)^{\otimes k}$ where $\Hil^1$ is the 2-dimensional single-qubit Hilbert space. Because $\Hil^k$ has the same dimension as $\Fock_k$ and a similar structure, they are isomorphic, and thus qubit operators can be bijectively mapped to fermionic operators and vice versa. This can be achieved using the Jordan-Wigner transform~\cite{jordan_uber_1928}. The transformations for the Majorana operators are:
\begin{equation}
    c_{2j-1} = \left(\prod_{k=1}^{j-1} Z_k \right) X_j, \quad c_{2j} = \left(\prod_{k=1}^{j-1} Z_k \right) Y_j,
\end{equation}
with X, Y, and Z being the standard Pauli operators.

\subsubsection{Parity}

The parity of a fermionic pure state is whether it contains an even number of fermions (even parity) or an odd number of fermions (odd parity). Fermionic mixed states must obey \textit{parity superselection}, that is, the density operator can only be a mixture of pure states with definite parity, i.e., pure states that are not superpositions of states of different parities. Such states are referred to as being even, though it should be noted that even states may mix pure states of definite even and definite odd parity. In addition, all physical unitaries acting on the state must preserve parity as internal dynamics cannot change parity.

Being a mixture of only definite parity pure states, i.e., parity eigenstates, even states are diagonalizable in the parity basis. Because all valid fermionic states must be even states, the Fock space for a system of fermions can be split into $\Fock_{k,+} \oplus \Fock_{k,-}$ where $+$ indicates the part with even parity (the even parity sector) and $-$ indicates the part with odd parity (the odd parity sector). Similarly, the density operator can be split into $\rho = \rho_+ \oplus \rho_-$ with a block diagonal form of
\begin{equation}
    \begin{bmatrix}
    \rho_{+} \hspace{5mm} & 0 \\
    0 \hspace{5mm} & \rho_{-}
    \end{bmatrix},
\end{equation}
where the diagonals represent the parts of the state with definite parity, and the off-diagonals represent the parts of the state with parity superpositions which must always be equal to zero for the state to obey parity superselection and thus be a valid fermionic state. Further, because such states are diagonalizable in the parity basis, they must commute with the \textit{parity operator} defined as
\begin{equation}
    P = (-1)^{\sum_i n_i},
\end{equation}
such that
\begin{equation}
    \begin{aligned}
        P \ket{even} &= \ket{even} \\
        P \ket{odd} &= -\ket{odd}.
    \end{aligned}
\end{equation}
This can be written via Jordan-Wigner as
\begin{equation}
    P = \prod_i Z_i.
\end{equation}

The action of $P$ on a pure state is
\begin{equation}
    P \ket{\psi} = \ket{even} - \ket{odd}.
\end{equation}
We can thus see that
\begin{equation}
    \ket{even} = \dfrac{1}{2}\left( \ket{\psi} + P \ket{\psi} \right), \quad \ket{odd} = \dfrac{1}{2}\left( \ket{\psi} - P \ket{\psi} \right).
\end{equation}
The operators used to project onto the even parity and odd parity subspaces are then
\begin{equation}
    \label{eq:parity_projection_operators}
    P_+ = \dfrac{1}{2} \left( I + P \right), \quad P_- = \dfrac{1}{2} \left( I - P \right),
\end{equation}
which are Hermitian and idempotent.

\subsubsection{Majorana Representation}

The set of all possible ordered products of the Majorana operators forms a complete orthonormal basis for all operators that act on $\Fock_k$, and thus any of these operators can be decomposed into sums of ordered products of the Majorana operators~\cite{bravyi_lagrangian_2004}. Let us define the ordered index set
\begin{equation}
    J = (i_1 < i_2 < \dots < i_l), \quad i_j \in \{1, \dots, 2k \},
\end{equation}
and let
\begin{equation}
    c_J = c_{i_1} c_{i_2} \dots c_{i_{|J|}}.
\end{equation}
Then, any operator $X$ can be written as
\begin{equation}
    X = \sum_{J \subseteq \{ 1, \dots, 2k \}} \alpha_J c_J.
\end{equation}
The coefficient $\alpha_J$ is a projection of $X$ onto $c_J$, and so is naturally the trace inner product $\Tr[X^\dagger c_J]$ multiplied by a normalization factor. This decomposition can be referred to as the \textit{Majorana decomposition} of $X$. Operators that preserve parity and respect parity superselection are sums of only even numbered products of the Majorana operators~\cite{melo_power_2013, oszmaniec_classical_2014}. The \textit{Majorana conjugate} of an operator $X$ can then be defined as
\begin{equation}
    \label{eq:majorana_conjugate}
    \Tilde{X} = \sum_{J \subseteq \{ 1, \dots, 2k \}} \alpha^*_J c_J.
\end{equation}
It is easy to see that the action of the Majorana conjugation operation is anti-unitary, that is, anti-linear (conjugate-linear) and inner-product preserving, as it complex conjugates the coefficients and acts as identity on the Majorana strings.

\subsubsection{Fermionic Gaussian}

Fermionic Gaussian states are even states that are ground and thermal states of systems of non-interacting (free) fermions~\cite{melo_power_2013}. They have quadratic non-interacting Hamiltonians and, by Wick's theorem, are fully characterized by their covariance matrix and are thus efficiently computable despite being able to maintain entanglement.

A quantum channel is pure Gaussian if it maps pure fermionic Gaussian input states to pure fermionic Gaussian output states. Similarly, a channel is fermionic convex Gaussian if it maps input states that are convex combinations of pure fermionic Gaussian states to output states that are convex combinations of pure fermionic Gaussian states. Such states are said to be in the \textit{convex hull} of fermionic Gaussians as the pure fermionic Gaussian states form a convex set in which they are the extreme points of the hull. All pure Gaussian channels are, by definition, convex Gaussian as well. Matchgates are the qubit analogues of fermionic Gaussian channels related by Jordan-Wigner.

\subsection{Noise}

There are a few different types of noise that a qubit can experience, but the most common and the type we shall use here is \textit{depolarizing noise}, in particular, \textit{local depolarizing noise}. Depolarizing noise randomizes (depolarizes) the state of a qubit, and can be represented by the following quantum channel~\cite{nielsen_chuang_2010}:
\begin{equation}
    \mathscr{D}_p(\rho) = (1-p) \rho + p \dfrac{I}{2},
\end{equation}
with p being the probability of depolarization and $0 \leq p \leq 1$, and $\rho$ being the density operator representing the qubit state. Being a completely positive quantum channel, the action of $\mathscr{D}_p(\rho)$ can be written in terms of Kraus operators as
\begin{equation}
    \mathscr{D}_p(\rho) = \sum_{i=0}^3 K_i \rho K_i^\dagger,
\end{equation}
with the Kraus operators:
\begin{equation}
    \begin{aligned}
        K_0 &= \sqrt{1-\dfrac{3p}{4}} I \\
        K_1 &= \sqrt{\dfrac{p}{4}} X \\
        K_2 &= \sqrt{\dfrac{p}{4}} Y \\
        K_3 &= \sqrt{\dfrac{p}{4}} Z.
    \end{aligned}
\end{equation}

Local depolarizing noise differs from global depolarizing noise in that it applies a depolarizing noise channel to each qubit with the same probability of depolarization in parallel rather than applying a single noise channel to every qubit. For two qubits, the local depolarizing noise channel has the Kraus form
\begin{equation}
    \label{eq:2channel_noise}
    \begin{aligned}
      \mathscr{D}^{\otimes 2}_p(\rho) &= (\mathscr{D}_p \otimes \mathscr{D}_p) (\rho) \\
        &= \sum_{i=0}^3 \sum_{j=0}^3 (K_i \otimes K_j) \rho (K_i \otimes K_j)^\dagger,
    \end{aligned}
\end{equation}
where $\rho$ is the density operator that represents the state of both qubits.

It is often more convenient to work with density matrices than with channels, and for such cases one can use the \textit{Choi state} representation of a quantum channel~\cite{nielsen_chuang_2010}. Such states are defined in $\Hil_S \otimes \Hil_R$ with $S$ being the system on which $\mathcal{E}$ acts and $R$ being a reference system of identical dimension. The Choi state is then given as
\begin{equation}
    \sigma = (\mathcal{E} \otimes I_R)(\ket{\alpha} \bra{\alpha}),
\end{equation}
with the joint state $\ket{\alpha} \in \Hil_S \otimes \Hil_R$ being a state that maximally entangles $S$ and $R$. This state fully characterizes the quantum channel, is unique to the channel, and allows analysis of its effect on any arbitrary state by determining its effect on a state maximally entangled with the reference. This correspondence is known as Choi-Jamiołkowski isomorphism~\cite{jamiolkowski_linear_transforms_1972, choi_linear_maps_1975}.

\begin{figure}[t]
  \centering
  \includegraphics[width=0.45\textwidth]{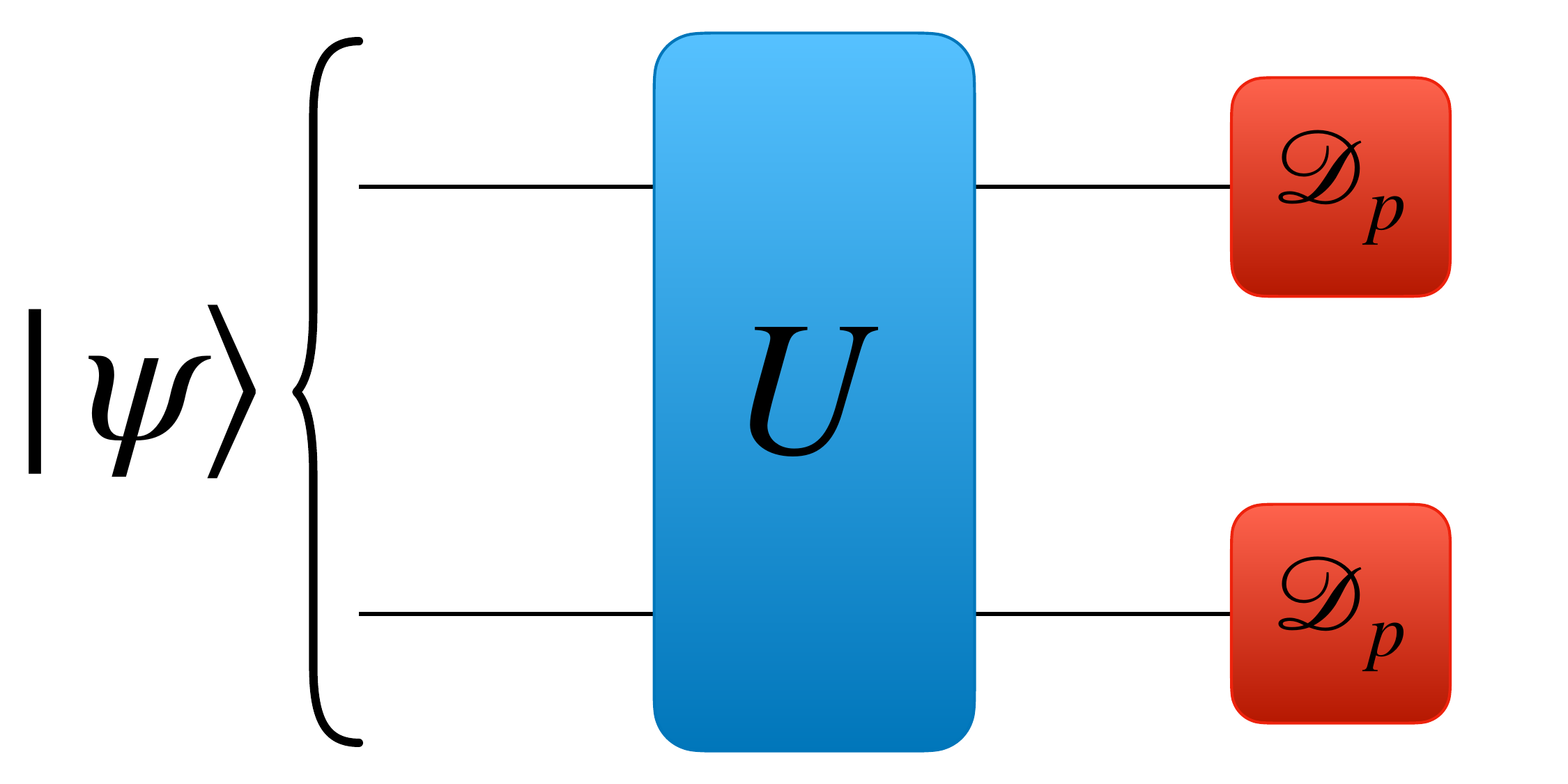}
  \caption{The noise model under consideration: First we apply the two qubit resourceful gate $U$, then apply local depolarizing noise $\mathscr{D}^{\otimes 2}_p$.} 
  \label{fig:noise_model}
\end{figure}

\section{Results}

In this section, we show that the fault-tolerance upper limit for fermionic quantum computing is approximately $62\%$ when using SWAP as a non-Gaussian resource. Given the non-Gaussian nature of SWAP~\cite{Jozsa_matchgates_simulation_2008, brod_matchgates_universal_2011}, one might expect this limit to be the highest among all 2-qubit parity preserving gates, and indeed, this is proven in Appendix~\ref{app:concurrence_reduction} and~\ref{app:concurrence_derivation} with a closed-form of $p=(8 - 2 \sqrt{6})/5$. In addition, we also show that this bound applies when using any maximally non-Gaussian gate such as SWAP or CZ as the resourceful gate. The basic idea is as follows. Consider a noisy circuit with matchgates and resourceful non-Gaussian gates where each of the latter is followed by two local depolarizing noise channels. If the noise is great enough, the quantum channel representing the non-Gaussian gate combined with the local depolarizing noise becomes convex Gaussian. In such a case, the circuit no longer has non-Gaussian support to extend it to universality and non-efficient computability, and thus fault-tolerant computation becomes effectively impossible. This is easy to see because all intermediary and output states would be confined to the convex hull of Gaussians instead of the entire set of possible states, and so cannot be universal. To determine if this is the case, we can start by finding the channel that describes the combination of our resourceful gate and the local depolarizing noise. From there, a fermionic Choi state that preserves convex Gaussianity can be found for the combined channel. Using this, we can determine if the Choi state and thus the combined channel is convex Gaussian by determining that it is an even state and that the fermionic analogs of the Uhlmann-Wootters concurrence for the Choi state are equal to zero~\cite{oszmaniec_classical_2014}. The probability of depolarization that first causes this to happen is the noise limit.

Let us refer to our 2-qubit parity preserving resourceful gate as $U$. Starting from the Kraus form of the 2-qubit local depolarizing noise in Eq.~\eqref{eq:2channel_noise}, we can get the combined channel with the 2-qubit unitary (visualized in Figure~\ref{fig:noise_model}) by first applying $U$,
\begin{equation}
    \begin{aligned}
        \mathcal{E}_{U,p}(\rho) &= (\mathscr{D}_p^{\otimes 2} \circ \mathscr{U} \otimes I)(\rho) \\
        &= \sum_{i=0}^3 \sum_{j=0}^3 (K_i \otimes K_j) U \rho U^\dagger (K_i \otimes K_j)^\dagger.
    \end{aligned}
\end{equation}
For brevity, let us define
\begin{equation}
    \label{eq:noise_kraus_simple}
    \sum_{i=0}^3 \sum_{j=0}^3 (K_i \otimes K_j) := \sum_\mu \sqrt{w_\mu} K_\mu,
\end{equation}
where we let $\mu = (i,j)$ and have $\sqrt{w_\mu} = \sqrt{w_i} \sqrt{w_j}$ be the multiplied coefficients from the Kraus operators and $K_{\mu=(i,j)}=K_{i}\otimes K_{j}$. Hiding the summation symbol for brevity, the combined channel can be written as
\begin{equation}
    \label{eq:combined_channel}
    \mathcal{E}_{U,p}(\rho) = w_\mu K_\mu U \rho U^\dagger K_\mu^\dagger.
\end{equation}

The corresponding Choi state for this quantum channel that preserves the convex Gaussian property can be derived using the maximally entangled Gaussian pure state defined as~\cite{bravyi_lagrangian_2004}
\begin{equation}\label{eq:max_entangled_fermion}
    \rho_\mathcal{E} = (\mathcal{E} \otimes_f I)(\rho_I),\quad \rho_I = \dfrac{1}{2^4} \prod_{a=1}^{4} \left( I + i c_a c_{4 + a} \right).
\end{equation}
For the case of two-qubit unitaries, expanding the Majorana monomials in $\rho_I$ introduces no reordering sign changes relative to the regular tensor product, so
\begin{equation}
    \rho_\mathcal{E} = (\mathcal{E} \otimes_f I)(\rho_I) = (\mathcal{E} \otimes I)(\rho_I).
\end{equation}
The state $\rho_{I}=\ketbra{\Phi_{G}}{\Phi_{G}}$ maximally entangles the modes between the base system (the first 2 modes) and the reference system (the last 2 modes). Therefore, $\ket{\Phi_G}$ and the standard Schmidt form for a 4-mode maximally entangled bipartite state $\ket{\Phi_4}$ are both purifications of the base system and thus differ only by a local unitary on the reference and can be written as given below~\cite{nielsen_chuang_2010}.
\begin{equation}
    \ket{\Phi_G} = (I \otimes R) \ket{\Phi_4}, \quad \ket{\Phi_4} = \frac{1}{2} \sum_x \ket{x}_S \ket{x}_R.
\end{equation}
The operator $R$ can then be determined by finding $\ket{\Phi_G}$ with the eigenvalue equations~\cite{bravyi_lagrangian_2004} $c_{4+a} \ket{\Phi_G} = -i c_a \ket{\Phi_G}$. With this, it is straightforward to find that an $R$ that satisfies these conditions acts as follows in the computational basis:
\begin{equation}
    \begin{aligned}
        R \ket{00} &= +1 \ket{11},\quad R \ket{01} = -i \ket{10} \\
        R \ket{10} &= +i \ket{01},\quad R \ket{11} = -1 \ket{00}.
    \end{aligned}
\end{equation}
Finally, we note that $\ket{\Phi_G}$ is a parity eigenstate of definite even parity, and so $P\ket{\Phi_G} = \ket{\Phi_G}$.

\begin{figure*}[t]
    \centering
    \includegraphics[width=0.95\linewidth]{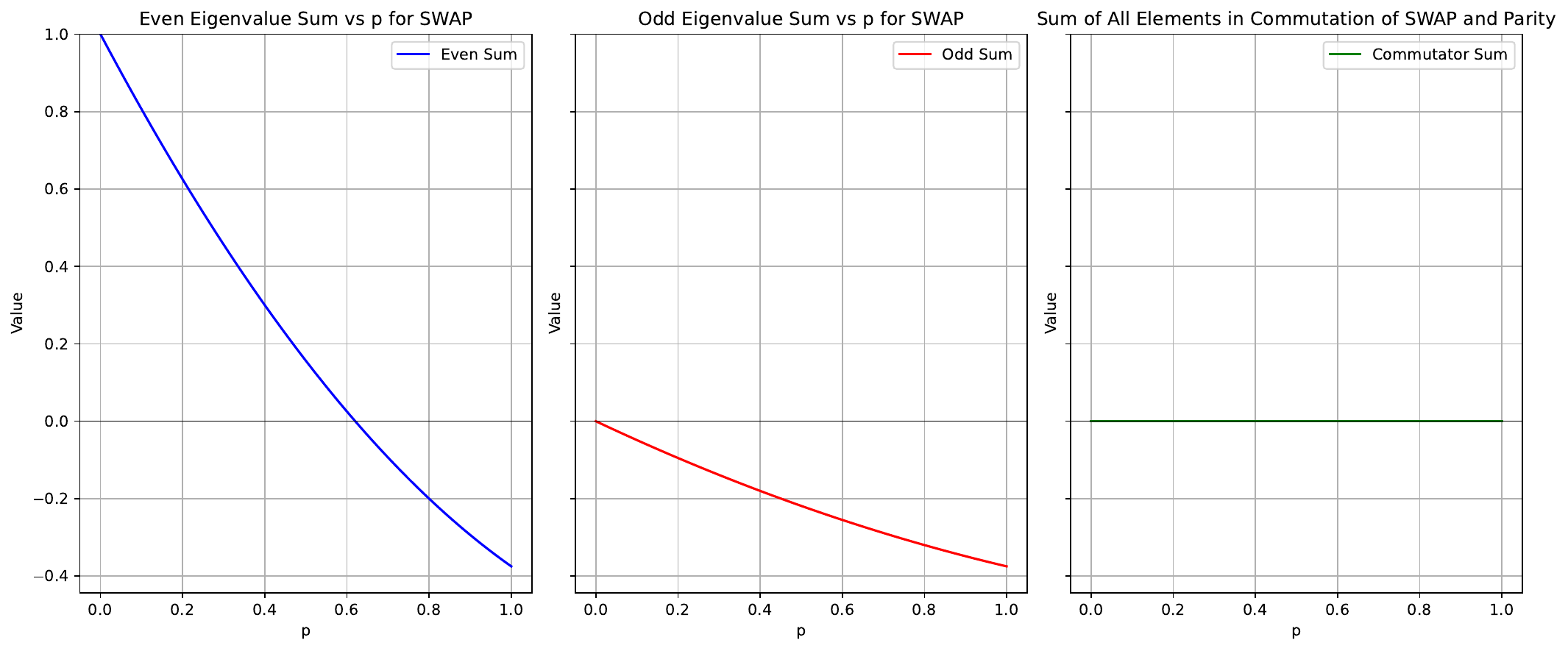}
    \caption{Numerical results implementing Algorithm \ref{alg:convex_gaussian_alg} for SWAP. The left two graphs show the value for $\Delta^\pm$ for different probabilities of depolarization $p$. The third graph shows that $\left[ \rho_{U,p}, P \right] = \boldsymbol{0}$ regardless of noise.}
    \label{fig:swap_graph}
\end{figure*}
Combining these expressions we find the total fermionic Choi state for Eq.~\eqref{eq:combined_channel} to be
\begin{equation} 
    \label{eq:fermionic_choi}
    \rho_{U,p} = w_\mu (K_\mu U \otimes I) \ket{\Phi_G} \bra{\Phi_G} (K_\mu U \otimes I)^\dagger.
\end{equation}
It is easy to see that applying the parity operator to this state always returns the same state, and so it always commutes with parity and is always a valid fermionic state given $U$ is a parity preserving gate. We can then determine if this Choi state is convex Gaussian to determine if the combined channel it corresponds to is also convex Gaussian. This can be done using the Uhlmann-Wootters concurrences for its even and odd parity projections~\cite{oszmaniec_classical_2014}. When these concurrences are equal to zero, the state lies within the convex hull of Gaussians and is thus convex Gaussian. These concurrences are defined for 4-mode fermionic states as
\begin{equation} 
    \label{eq:concurrence}
    C_\pm(U,p) = C_\pm(\rho_{U,p}) = \max \left( 0, \lambda_1^\pm - \sum_{k=2}^8 \lambda_k^\pm \right).
\end{equation}
For convenience, we will define the symbol
\begin{equation}
    \Delta^\pm = \lambda_1^\pm - \sum_{k=2}^8 \lambda_k^\pm.
\end{equation}
Here, $(\lambda_1^\pm, \lambda_2^\pm, \dots)$ is the non-increasing (largest to smallest) set of the singular values of $\sqrt{\rho_\pm} \sqrt{\widetilde{\rho_\pm}}$, or the positive square roots of the eigenvalues of the operator $\sqrt{\rho_\pm} \widetilde{\rho_\pm} \sqrt{\rho_\pm}$~\cite{oszmaniec_classical_2014, uhlmann_concurrence}. These are also equivalent to the positive square roots of the eigenvalues of $\rho_\pm \widetilde{\rho_\pm}$ as it only differs from $\sqrt{\rho_\pm} \widetilde{\rho_\pm} \sqrt{\rho_\pm}$ by a two-matrix swap and thus shares the same eigenvalues. Here, $\rho_\pm = P_\pm \rho P_\pm$ are the parity subspace projections of $\rho$, and $\widetilde{\rho_\pm}$ is the Majorana conjugate of $\rho_\pm$ defined in Eq.~\eqref{eq:majorana_conjugate}. 

Because the channel is convex Gaussian if its Choi state formed in Eq.~\eqref{eq:fermionic_choi} is convex Gaussian,
\begin{equation}
    \boxed{\mathcal{E}_{U,p} \; \text{ is convex Gaussian} \Longleftrightarrow C_+(U,p) = C_-(U,p) = 0.}
\end{equation}
Using this gives a complete means by which one can determine whether the combined quantum channel of the 2-qubit parity preserving unitary $U$ and the local depolarizing noise is convex Gaussian. This is summarized in Algorithm~\ref{alg:convex_gaussian_alg}, and the numerical results for SWAP swept across all values of $p$ are shown in Figure~\ref{fig:swap_graph}.

\begin{algorithm}
    \label{alg:convex_gaussian_alg}
    \caption{Determining Convex Gaussianity for 2-Qubit Parity Preserving Unitary with Noise}
    $U \gets \text{2-qubit parity preserving unitary}$ \;
    $\mathcal{E}_{U,p} \gets \text{U with local depolarizing noise}$ \;
    $\rho_{U,p} \gets \text{Fermionic Choi state for } \mathcal{E}_{U,p}$ \;
     \eIf{$C_+(\rho_{U,p}) = C_-(\rho_{U,p}) = 0$}
     {$\mathcal{E}_{U,p}$ is convex Gaussian \;}{$\mathcal{E}_{U,p}$ is not convex Gaussian \;}
\end{algorithm}

Analytically, the positive concurrence of the combined channel reduces to a closed form that depends on the non-Gaussianity of $U$ through its block determinants. For a parity preserving unitary $U = A \oplus B$, with $A$ and $B$ its even- and odd-parity $2\times2$ blocks, we define the determinant separation $\eta(U) = \tfrac{1}{2}|\det A - \det B| \in [0,1]$. This quantity vanishes for Gaussian (matchgate) gates and equals 1 for highly non-Gaussian gates such as SWAP and CZ. The odd-parity concurrence is always zero and the even-parity concurrence is
\begin{equation}
    C_+(U,p) = \max\left(0, \; \tfrac{1}{2}\eta \left(1-p\right)\left(2-p\right) + \tfrac{p^2}{8} - \tfrac{p}{2}\right),
\end{equation}
as derived in Appendix~\ref{app:concurrence_reduction} and~\ref{app:concurrence_derivation}. All parity preserving 2-qubit gates $U$ can be decomposed into a Gaussian part and a non-Gaussian part that is parameterized by $\eta$. The Gaussian part does not affect the eigenvalues of $\rho_\pm \widetilde{\rho_\pm}$, and so does not affect the values of the concurrences whose dependence on $U$ is thus only the parameter $\eta$. The eigenvalues of $\rho_\pm \widetilde{\rho_\pm}$ are equivalent to the singular values of an $8\times8$ matrix that is block diagonalizable over the 16 even and odd parity orthogonal noise branches formed with $K_\mu$. Since the positive concurrence increases with $\eta$, the noise limit is maximized over all 2-qubit parity preserving gates by taking $\eta=1$ which sets the positive concurrence to zero at
\begin{equation}
    p^+_\text{max} = \dfrac{8 - 2 \sqrt{6}}{5} \approx 0.6202,
\end{equation}
achieved by any gate whose block determinants are maximally far apart such as SWAP and CZ. Because this limit is set by the point at which every such gate under noise loses its resourcefulness, the bound holds independently of circuit depth.

\section{Conclusion}

In this paper, we studied the noise limits of fault-tolerant computation under local depolarizing noise for circuits consisting of matchgates and non-Gaussian resource gates, and thus for fermionic quantum computation. We found this noise limit by determining when the fermionic Choi state for a two-qubit parity preserving unitary combined with local depolarizing noise becomes fermionic convex Gaussian. By using Uhlmann-Wootters concurrences, we were able to find the level of noise at which this occurs to be $p=(8-2\sqrt{6})/5$ or approximately $62~\%$. We also prove that this point is maximized for 2-qubit parity preserving gates such as SWAP and CZ whose block determinants are maximally far apart, that is, they are maximally non-Gaussian. Therefore, $\approx62~\%$ serves as the upper bound limit for local depolarizing noise in fermionic quantum computing. Because we analyzed the point at which every single 2-qubit parity preserving unitary under noise loses its resourcefulness, this bound applies regardless of the depth of the circuit.

Interestingly, the noise limit of $\approx 0.62$ is higher than the one found in Ref.~\cite{buhrman_new_2006} for Clifford + T at $\approx 0.45$. As discussed previously, the bound found with Clifford gates and T is not directly comparable because reconstructing matchgate circuits with non-Gaussian resourceful gates using Clifford and T gates results in a noise model different from the one used to get the $\approx 0.45$ limit. Further, in fermionic quantum computation, valid gates must be parity preserving to obey superselection rules, so the Clifford + T gate set cannot be physically implemented. The $p \approx 62~\%$ limit is thus the best depth-independent limit currently known for fermionic quantum computing to the authors' knowledge.

\section{Future Work}

The method described in this paper is only applicable to 2-qubit parity preserving gates as only parity preserving gates will correspond to valid fermionic Choi states, and only 2-qubit gates will correspond to 4-mode Choi states with which the concurrence method applies. Further, we only characterize the resourcefulness of the combined gate and noise channel when it comes to resourcefulness in matchgate circuits. Therefore, future work may seek to generalize the method described here to arbitrary n-qubit unitary gates and to other resource theories. For instance, a PPT-like condition has recently been identified for fermions~\cite{garcia_fermionic_2026}. This could include studying arbitrary two-qubit channels, replacing local depolarizing noise by more general noise models, and determining thresholds for resources such as entanglement or spin coherence; such thresholds could connect directly to applications including quantum networks and quantum sensing. One could ask whether criteria analogous to the concurrences used here exist that can be used to characterize other resources, and whether the computation thereof would similarly reduce to tractable calculation. These remain open and interesting directions.

Because the techniques used for the Clifford bound in Ref.~\cite{buhrman_new_2006} are different from the ones presented here, it would be interesting to repeat this analysis in that setting to get a more direct comparison of these results to the Clifford case so as to determine if one gate set is more noise resistant than the other. Namely, consider a general two qubit gate and find the local depolarizing noise needed to map its Choi state into the convex hull of stabilizer states. Unlike the fermionic convex Gaussian case, it is unclear how to determine if a mixed state is a convex combination of stabilizer states (i.e. it lacks magic) efficiently. In particular the robustness of magic~\cite{howard_application_2017} is a natural choice and can be computed using linear programming, however the parameterization of gates makes finding an optimum potentially challenging. Additionally, one would need to compare the bound found in this work in terms of fermionic concurrences to a robustness of fermionic magic.

\section{Acknowledgments}
The authors thank Xun Gao for helpful guidance during the creation of this work. L.C. acknowledges support from the National Science Foundation Graduate Research Fellowship under Grant No. 2140743 and the QSE G1 Fellowship at Harvard. Any opinions, findings, and conclusions or recommendations expressed in this material are those of the author(s) and do not necessarily reflect the views of the National Science Foundation.

\appendix

\section{Concurrence Parameterization}
\label{app:concurrence_reduction}

Every 2-qubit parity preserving unitary gate can be written in the computational basis as~\cite{brod_matchgates_universal_2011}
\begin{equation}
    U = G(A, B) = 
    \begin{bmatrix}
    A_{11} & 0 & 0 & A_{12} \\
    0 & B_{11} & B_{12} & 0 \\
    0 & B_{21} & B_{22} & 0 \\
    A_{21} & 0 & 0 & A_{22}
    \end{bmatrix},
\end{equation}
where
\begin{equation}
    A, B \in U(2).
\end{equation}
The matrix blocks, being unitaries, by definition must satisfy the constraint that $|\det A| = |\det B| = 1$. When $\det A = \det B$ exactly, $U$ is matchgate and fermionic Gaussian. To characterize these parity preserving unitaries, we can define the quantity
\begin{equation}
    \eta(U) = \frac{1}{2} | \det A - \det B | \in [0,1].
\end{equation}

Noticing again the magnitudes of $\det A$ and $\det B$ are equal to 1, we can define a quantity that lies on the unit circle parameterized by the phase difference of the determinants of A and B, $\delta$, as
\begin{equation}
    e^{i \delta} = \dfrac{\det A}{\det B} \in U(1).
\end{equation}
We can parameterize $\eta$ in terms of $\delta$ as follows
\begin{equation}
    \begin{aligned}
        \eta(U) &= \frac{1}{2} | \det A - \det B | \\
        &= \frac{1}{2} |\det B| \cdot |e^{i \delta} - 1| \\
        &= \frac{1}{2} |e^{i \delta} - 1| \\
        &= \left \vert\sin{\frac{\delta}{2}}\right \vert.
    \end{aligned}
\end{equation}

With this, one might hope to write $U$ in terms of a Gaussian part and non-Gaussian part parameterized by $\delta$. To do this, let us define the non-Gaussian part as
\begin{equation}
    V_\delta = \exp(i \delta P / 4),
\end{equation}
which in the computational basis has the form
\begin{equation}
    V_\delta = \text{diag}(e^{i \delta / 4}, e^{-i \delta / 4}, e^{-i \delta / 4}, e^{i \delta / 4}).
\end{equation}
One might then expect the Gaussian part to be
\begin{equation}
    G = V_\delta^\dagger U.
\end{equation}
Multiplying this out results in
\begin{equation}
    G_A = e^{-i \delta /4} A, \quad G_B = e^{i\delta/4} B.
\end{equation}
Thus the determinants of the A and B blocks are
\begin{equation}
    \det G_A = e^{-i \delta /2} \det A, \quad \det G_B = e^{i\delta/2} \det B,
\end{equation}
and their ratio is
\begin{equation}
    \begin{aligned}
        \dfrac{\det G_A}{\det G_B} &= e^{-i \delta} \dfrac{\det A}{\det B} \\
        &= e^{-i \delta} e^{i \delta} \\
        &= 1.
    \end{aligned}
\end{equation}
The operator $G$ is therefore a matchgate and Gaussian. We can then write any 2-qubit parity preserving unitary in terms of a non-matchgate component parameterized by $\delta$ and matchgate component as
\begin{equation}
    \label{eq:unitary_decomp}
    U = V_\delta G.
\end{equation}

Let us now apply a Gaussian unitary $G$ to the maximally entangled Gaussian state from Eq.~\eqref{eq:max_entangled_fermion}. We get
\begin{equation}
    \begin{aligned}
        (G \otimes I) \ket{\Phi_G} &= (G \otimes R) \ket{\Phi_4} \\
        &= (I \otimes R) (G \otimes I) \ket{\Phi_4}.
    \end{aligned}
\end{equation}
Using the standard transfer identity for maximally entangled bipartite states~\cite{nielsen_chuang_2010}, this becomes
\begin{equation}
    \begin{aligned}
        (G \otimes I) \ket{\Phi_G} &= (I \otimes R) (I \otimes G^T) \ket{\Phi_4} \\
        &= (I \otimes R G^T) \ket{\Phi_4} \\
        &= (I \otimes R G^T R^\dagger) \ket{\Phi_G} \\
        &= (I \otimes \hat{G}) \ket{\Phi_G}.
    \end{aligned}
\end{equation}
Transposition and unitary action leave the determinants of the matrix blocks A and B unchanged, so $\hat{G}$ remains Gaussian.

Let us again return to our combined channel Choi state from Eq.~\eqref{eq:fermionic_choi}. Using the decomposition in Eq.~\eqref{eq:unitary_decomp}, this can be written as
\begin{equation}
    \begin{aligned}
        \rho_{U,p} &= w_\mu (K_\mu V_\delta G \otimes I) \ket{\Phi_G} \bra{\Phi_G} (K_\mu V_\delta G \otimes I)^\dagger \\
        &= w_\mu (K_\mu V_\delta \otimes \hat{G}) \ket{\Phi_G} \bra{\Phi_G} (K_\mu V_\delta \otimes \hat{G})^\dagger \\
        &= (I \otimes \hat{G}) \rho_{V_\delta,p} (I \otimes \hat{G})^\dagger.
    \end{aligned}
\end{equation}
To make this more compact, let's define the unitary Gaussian operator $\Gamma$ such that
\begin{equation}
    \rho_{U,p} = \Gamma \rho_{V_\delta,p} \Gamma^\dagger.
\end{equation}

The concurrences that we're interested in are dependent on the positive square roots of the eigenvalues of the operator $\rho_\pm \widetilde{\rho_\pm}$. Let's examine how the action of $\Gamma$ affects this operator. We can see that, Gaussian operators being physical even states, $\Gamma$ must commute with the parity operator and thus the parity projection operators. Therefore,
\begin{equation}
    \begin{aligned}
        (\rho_{U,p})_\pm &= P_\pm \Gamma \rho_{V_\delta,p} \Gamma^\dagger P_\pm \\
        &= \Gamma (\rho_{V_\delta,p})_\pm \Gamma^\dagger.
    \end{aligned}
\end{equation}
Further, the action of a Gaussian unitary on Majorana operators is represented by a real matrix~\cite{bravyi_lagrangian_2004}, and so taking the complex conjugate of the coefficients in the Majorana expansion before or after applying $\Gamma$ makes no difference. Thus,
\begin{equation}
    \widetilde{\Gamma X \Gamma^\dagger} = \Gamma \tilde{X} \Gamma^\dagger,
\end{equation}
and
\begin{equation}
    \begin{aligned}
        \widetilde{(\rho_{U,p})_\pm} & = \widetilde{\Gamma (\rho_{V_\delta,p})_\pm \Gamma^\dagger} \\
        &= \Gamma \widetilde{(\rho_{V_\delta,p})_\pm} \Gamma^\dagger.
    \end{aligned}
\end{equation}
Finally,
\begin{equation}
    \begin{aligned}
        (\rho_{U,p})_\pm \widetilde{(\rho_{U,p})_\pm} &= \Gamma (\rho_{V_\delta,p})_\pm \Gamma^\dagger \Gamma \widetilde{(\rho_{V_\delta,p})_\pm} \Gamma^\dagger \\
        &= \Gamma (\rho_{V_\delta,p})_\pm \widetilde{(\rho_{V_\delta,p})_\pm} \Gamma^\dagger.
    \end{aligned}
\end{equation}
We can see that this a similarity transform of $(\rho_{V_\delta,p})_\pm \widetilde{(\rho_{V_\delta,p})_\pm}$ and thus they share the same eigenvalues. Therefore,
\begin{equation}
     \boxed{C_\pm(U, p) = C_\pm(V_\delta, p) \text{ for all } p.\ }
\end{equation}

\section{Derivation of Concurrences}
\label{app:concurrence_derivation}

For brevity, let us define
\begin{equation}
    \ket{\psi} := (V_\delta \otimes I) \ket{\Phi_G}.
\end{equation}
The parameterized noisy fermionic Choi state can then be written as
\begin{equation}
    \label{eq:parameterized_choi}
    \begin{aligned}
        \rho_{V_\delta,p} &= w_\mu (K_\mu \otimes I) \ket{\psi} \bra{\psi} (K_\mu \otimes I)^\dagger \\
        &= w_\mu \ket{\psi_\mu} \bra{\psi_\mu}.
    \end{aligned}
\end{equation}
The vectors $\ket{\psi}$ and $\ket{\psi_\mu}$ are created by applying only local unitaries to $\ket{\Phi_G}$, and so remain maximally entangled. Therefore,
\begin{equation}
    \begin{aligned}
        \braket{\psi_\mu}{\psi_\nu} &= \bra{\psi} (K_\mu \otimes I)^\dagger (K_\nu \otimes I) \ket{\psi} \\
        &= \bra{\psi} (K_\mu K_\nu \otimes I) \ket{\psi} \\
        &= \text{Tr} \left[ (K_\mu K_\nu \otimes I) \ketbra{\psi}{\psi} \right] \\
        &= \text{Tr}_S \left[ K_\mu K_\nu \text{Tr}_R \ketbra{\psi}{\psi} \right] \\
        &= \frac{1}{4} \text{Tr} \left[ K_\mu K_\nu \right].
    \end{aligned}
\end{equation}
The trace of a Pauli string is zero unless it is proportional to the identity operator, so
\begin{equation}
    \text{Tr} \left[ K_\mu K_\nu \right] = 4 \delta_{\mu \nu},
\end{equation}
and thus
\begin{equation}
    \begin{aligned}
        \braket{\psi_\mu}{\psi_\nu} &= \delta_{\mu \nu}.
    \end{aligned}
\end{equation}
Therefore, the 16 noise branches of $\ket{\psi}$ form an orthonormal eigenbasis for the noisy fermionic Choi state and its eigenvalues are the values of $w_\mu$. The eigenvalue spectrum of the Choi state is thus
\begin{equation}
    \label{eq:eigenval_spectrum}
    \left\{ \left( 1 - \frac{3p}{4} \right)^2 \; (\times1), \quad \dfrac{4p - 3p^2}{16} \; (\times6), \quad \frac{p^2}{16} \; (\times9) \right\}.
\end{equation}

We can see that $\ket{\psi_\mu}$ is a $\pm1$ parity eigenstate as
\begin{equation}
    \begin{aligned} 
        P \ket{\psi_\mu} &= (P_S K_\mu \otimes P_R) \ket{\psi} \\
        &= \pm (K_\mu \otimes I) P \ket{\psi} \\
        &= \pm \ket{\psi_\mu},
    \end{aligned}
\end{equation}
where the sign depends on whether the Pauli string $K_\mu$ commutes or anti-commutes with the parity operator. The strings that commute are
\begin{equation}
    II, \; IZ, \; XX, \; XY, \; YX, \; YY, \; ZI, \; ZZ,
\end{equation}
and the strings that anti-commute are
\begin{equation}
    IX, \; IY, \; XI, \; XZ, \; YI, \; YZ, \; ZX, \; ZY.
\end{equation}
The even parity eigenvectors of the fermionic Choi state are then the $\ket{\psi_\mu}$ that correspond to the strings that commute, and the odd parity eigenvectors are the ones that correspond to the strings that anti-commute.

We can see that the even and odd parity projected states can be written as
\begin{equation}
    \rho_\pm = \sum_{\mu \in \pm} w_\mu \ket{\psi_\mu} \bra{\psi_\mu}.
\end{equation}
We can define the matrix $A_\pm$ as a $16 \times 8$ matrix whose eight columns are the unnormalized eigenvectors $\sqrt{w_\mu} \ket{\psi_\mu}$ in the corresponding parity sector. Then,
\begin{equation}
    \rho_\pm = A_\pm A_\pm^\dagger.
\end{equation}
To get a matrix for the Majorana conjugation of this, we need a matrix representation of the operation. Let
\begin{equation}
    \Theta = JC,
\end{equation}
be the anti-unitary that implements Majorana conjugation. We want the action of $\Theta$ to complex-conjugate coefficients of Majorana strings without altering the strings themselves. Thus, $C$ implements ordinary complex conjugation and $J$ compensates for any changes to the Majorana operators introduced from complex conjugation. One can convince themselves that the even-indexed Majoranas are precisely the factors needed for this compensation by noticing that $J$ must commute with the odd-indexed Majoranas and anti-commute with the even-indexed Majoranas, and so
\begin{equation}
    J := c_2 c_4 c_6 c_8 = -(X \otimes Y)_S \otimes (X \otimes Y)_R.
\end{equation}
We can compact this to
\begin{equation}
    J = -\mathcal{J}_S \otimes \mathcal{J}_R,
\end{equation}
with
\begin{equation}\label{eqn:conjugation}
    \mathcal{J} := X \otimes Y.
\end{equation}

Writing out the Majorana conjugation of $\rho_\pm$ with these matrices, we get
\begin{equation}
    \begin{aligned}
        \widetilde{\rho_\pm} &= JC (A_\pm A_\pm^\dagger) C^\dagger J^\dagger \\
        &= J (A_\pm A_\pm^\dagger)^* J^\dagger \\
        &= J A_\pm^* A_\pm^T J^\dagger
    \end{aligned}
\end{equation}
We can then define
\begin{equation}
    B_\pm := J A_\pm^*.
\end{equation}
Then,
\begin{equation}
    \widetilde{\rho_\pm} = B_\pm B_\pm^\dagger.
\end{equation}
We can then write $\rho_\pm \widetilde{\rho_\pm}$ as
\begin{equation}
    \rho_\pm \widetilde{\rho_\pm} = A_\pm A_\pm^\dagger B_\pm B_\pm^\dagger.
\end{equation}
A two-matrix reordering can be done without changing the eigenvalues such that $\rho_\pm \widetilde{\rho_\pm}$ shares its eigenvalues with the operator
\begin{equation}
    (A_\pm^\dagger B_\pm) (A_\pm^\dagger B_\pm)^\dagger = \tau_\pm \tau_\pm^\dagger,
\end{equation}
where we define
\begin{equation}
    \tau_\pm := A_\pm^\dagger B_\pm = A_\pm^\dagger J A_\pm^*,
\end{equation}
and is an $8 \times 8$ matrix. By definition, the singular values of $\tau_\pm$ are the positive square roots of the eigenvalues of $\tau_\pm \tau_\pm^\dagger$ and thus $\rho_\pm \widetilde{\rho_\pm}$, the exact values we need for computing the concurrences. Thus, we can determine the concurrences by finding the singular values of $\tau_\pm$.

Writing out the terms of the matrix $\tau_\pm$ explicitly, we get
\begin{equation}
    \begin{aligned}
        (\tau_\pm)_{\mu \nu} &= (\sqrt{w_\mu} \bra{\psi_\mu}) J(\sqrt{w_\nu} \ket{\psi_\nu})^* \\
          &= \sqrt{w_\mu w_\nu} \bra{\psi_\mu} J \ket{\psi_\nu}^*,
    \end{aligned}
\end{equation}
which we can simplify to
\begin{equation}
    (\tau_\pm)_{\mu \nu} = \sqrt{w_\mu w_\nu} \; \Omega_{\mu \nu},
\end{equation}
where we define
\begin{equation}
    \label{eq:omega}
    \Omega_{\mu \nu} := \bra{\psi_\mu} J \ket{\psi_\nu}^* = \bra{\psi_\mu} \Theta \ket{\psi_\nu}.
\end{equation}
We can see that $\Omega$ is just the matrix representation of $\Theta$ in the basis formed by the noise branches of $\ket{\psi}$. By expanding this, we get
\begin{equation}
    \label{eq:omega_expanded}
    \begin{aligned}
        \Omega_{\mu \nu} &= \bra{\psi} (K_\mu \otimes I) (-\mathcal{J} \otimes \mathcal{J}) (K_\nu^* \otimes I) \ket{\psi}^* \\
        &= - \bra{\psi} (K_\mu \mathcal{J} K_\nu^* \otimes \mathcal{J}) \ket{\psi}^*,
    \end{aligned}
\end{equation}
\clearpage
Because $\ket{\psi}$ is maximally entangled, the transfer identity applies and so there exists an operator $W$ such that
\begin{equation}
    \begin{aligned}
        \Omega_{\mu \nu} &= - \bra{\psi} (K_\mu \mathcal{J} K_\nu^* W \otimes I) \ket{\psi} \\
        &= - \frac{1}{4} \text{Tr} \left[ K_\mu \mathcal{J} K_\nu^* W \right].
    \end{aligned}
\end{equation}
It is straightforward to show (as seen in Appendix~\ref{app:w_matrix_derivation}) that $W$ is
\begin{equation}
    W = \mathcal{J} \left( i \sin{\frac{\delta}{2}} - \cos{\frac{\delta}{2}} P \right).
\end{equation}
Then,
\begin{equation}
    \Omega_{\mu \nu} = -\frac{1}{4} \left( i \sin{\frac{\delta}{2}} \text{Tr} \left[ K_\mu \mathcal{J} K_\nu^* \mathcal{J} \right] - \cos{\frac{\delta}{2}} \text{Tr} \left[ K_\mu \mathcal{J} K_\nu^* \mathcal{J} P \right]  \right).
\end{equation}

We can now see that there are two conditions in which $\Omega_{\mu \nu}$ doesn't vanish. Firstly, using the properties of Pauli strings again, we can see that either $K_\mu \mathcal{J} K_\nu^* \propto \mathcal{J}$ such that $ K_\mu \mathcal{J} K_\nu^* \mathcal{J} \propto I$ or $K_\mu \mathcal{J} K_\nu^* \propto \mathcal{J} P$ such that $ K_\mu \mathcal{J} K_\nu^* \mathcal{J}P \propto I$, otherwise the traces vanish to zero. These break down into four single-mode conditions:
\begin{equation}
    \begin{aligned}
        K_{\mu,1} X K_{\nu,1}^* &\propto X, \quad  K_{\mu,2} K_{\nu,2}^* \propto Y \\
        K_{\mu,1} X K_{\nu,1}^* &\propto Y, \quad K_{\mu,2} Y K_{\nu,2}^* \propto X.
    \end{aligned}
\end{equation}
One can see that the first two conditions necessitate that $K_\nu = K_\mu$ to keep the first term and the second two conditions necessitate $K_\nu = K_\mu P$ to keep the second, and that these couple diagonal elements in $\Omega$ to non-diagonals in the same parity sector (P doesn't affect the parity sectors of $K_\mu$). For example, the diagonal element $\Omega_{(II) (II)}$ has a corresponding off-diagonal element at $\Omega_{(II) (ZZ)}$, and the same for $\Omega_{(ZZ) (ZZ)}$ and $\Omega_{(ZZ) (II)}$. The even parity sector indices are coupled as
\begin{equation}
    \{II,ZZ\}, \; \{IZ,ZI\}, \; \{XX,YY\}, \; \{XY,YX\},
\end{equation}
and the odd parity sector indices as
\begin{equation}
    \{IX,ZY\}, \; \{IY,ZX\}, \; \{XI,YZ\}, \; \{XZ,YI\}.
\end{equation}
Here, we're using $K_\mu$ and $K_\nu$ to represent their corresponding indices. Additionally, because we haven't introduced any coupling between parity sectors, $\Omega$ is block diagonalizable with the coupled diagonal and off-diagonal entries forming $2\times2$ matrix blocks. The magnitudes of all of the elements in $\Omega_{\mu \nu}$ are then
\begin{equation}
         |\Omega_{(K) (K)}| = \frac{1}{4} \left|4i \sin{\frac{\delta}{2}}\right| = \eta,
\end{equation}
and
\begin{equation}
    |\Omega_{(K) (KP)}| = \frac{1}{4} \left|4\cos{\frac{\delta}{2}}\right| = \sqrt{1 - \eta^2}.
\end{equation}

The eight total blocks in $\tau_+$ and $\tau_-$ can be written generically as
\begin{equation}
    \tau^{(B)} =
    \begin{bmatrix}
        w_1 \, \Omega^{(B)}_{11} & \sqrt{w_1 w_2} \ \Omega^{(B)}_{12} \\
        \sqrt{w_1 w_2} \; \ \Omega^{(B)}_{21} & w_2 \, \Omega^{(B)}_{22}
    \end{bmatrix},
\end{equation}
where $\sqrt{w_1}$ and $\sqrt{w_2}$ are the coefficients from Eq.~\eqref{eq:noise_kraus_simple} corresponding to the two coupled indices in the block. We now must find the two singular values of this block, $s_+ \geq s_-$. Let us start from a generic polynomial with roots $s_+$ and $s_-$,
\begin{equation}
    (x-s_+)(x-s_-) = x^2 - (s_+ + s_-)x + s_+ s_-.
\end{equation}
Using the quadratic formula, we get
\begin{equation}
    \begin{aligned}
        s_\pm &= \frac{1}{2} \left[ (s_+ + s_-) \pm \sqrt{(s_+ + s_-)^2 - 4 s_+ s_-} \right] \\
        &= \frac{1}{2} \left[ (s_+ + s_-) \pm \sqrt{(s_+ - s_-)^2} \right] \\
        &= \frac{1}{2} \left[ (s_+ + s_-) \pm (s_+ - s_-) \right],
    \end{aligned}
\end{equation}
and from rearranging the discriminant,
\begin{equation}
    s_+ + s_- = \sqrt{(s_+ - s_-)^2 + 4 s_+ s_-}.
\end{equation}
We can see that we can get $s_\pm$ from knowing $(s_+ - s_-)$ and $(s_+ s_-)$.

Being a matrix representation of the anti-unitary operator $\Theta$, the $\Omega$ matrix is unitary. Additionally, because $\Omega$ is block diagonal, all of its blocks are unitary as well and have determinant with magnitude of 1. Therefore,
\begin{equation}
    s_+ s_- = |\det \tau^{(B)}| = w_1 w_2 |\det \Omega^{(B)}| = w_1 w_2.
\end{equation}

The trace operator gives the sum of the eigenvalues of a matrix. The eigenvalues of $\tau^{(B)} \tau^{(B)\dagger}$ are equal to the squares of the singular values of $\tau^{(B)}$ by definition. Thus,
\begin{equation}
    \begin{aligned}
        s_+^2 + s_-^2 &= \text{Tr} \left[ \tau^{(B)} \tau^{(B)\dagger} \right]= \sum_{i,j} |\tau^{(B)}_{ij}|^2 \\
        &= \eta^2 w_1^2 + \eta^2 w_2^2 + 2 w_1 w_2 (1 - \eta^2).
    \end{aligned}
\end{equation}
Then,
\begin{equation}
    \begin{aligned}
        (s_+ - s_-)^2 &= s_+^2 + s_-^2 - 2 s_+ s_- \\
        &= \eta^2 (w_1^2 + w_2^2) - 2 w_1 w_2 \eta^2 \\
        &= \eta^2 (w_1 - w_2)^2.
    \end{aligned}
\end{equation}

With all this, the singular values for each block are
\begin{equation}
    \label{eq:singular_values}
    s_\pm = \frac{1}{2} \left( \sqrt{\eta^2 (w_1 - w_2)^2 + 4 w_1 w_2} \: \pm \eta |w_1 - w_2| \right).
\end{equation}
We can notice that, firstly, the singular values are the same if $w_1$ is swapped for $w_2$ and vice versa, and that secondly, when $w_1 = w_2$, the singular values reduce to
\begin{equation}
    s_\pm = w_1 = w_2.
\end{equation}

The even pair blocks have the following $w_1, w_2$ values
\begin{equation}
    \begin{aligned}
        \{II, ZZ\} &\longrightarrow{} \left( \left(1-\frac{3p}{4}\right)^2, \quad \frac{p^2}{16} \right) \\
        \{IZ, ZI\} &\longrightarrow{} \left( \dfrac{4p - 3p^2}{16}, \quad \dfrac{4p - 3p^2}{16} \right) \\
        \{XX, YY\} &\longrightarrow{} \left( \frac{p^2}{16}, \quad \frac{p^2}{16} \right) \\
        \{XY, YX\} &\longrightarrow{} \left( \frac{p^2}{16}, \quad \frac{p^2}{16} \right).
    \end{aligned}
\end{equation}
Plugging these into Eq.~\eqref{eq:singular_values} and simplifying results in the following singular values:
\begin{equation}
    \begin{aligned}
        \sqrt{\frac{1}{16} u^2 + \left(\frac{4p - 3p^2}{16}\right)^2} + \frac{1}{4} u, \quad (\times1) \\
        \sqrt{\frac{1}{16} u^2 + \left(\frac{4p - 3p^2}{16}\right)^2} - \frac{1}{4} u, \quad (\times1) \\
        \frac{4p - 3p^2}{16}, \quad (\times2) \\
        \frac{p^2}{16}, \quad (\times4),
    \end{aligned}
\end{equation}
with $u = \eta \left(1-p\right)\left(2-p\right)$.
It's simple to see that the first singular value is the largest. Thus,
\begin{equation}
    \label{eq:positive_delta}
    \Delta^+ = \frac{1}{2}\eta \left(1-p\right)\left(2-p\right) + \frac{p^2}{8} - \frac{p}{2}.
\end{equation}
The positive concurrence for the fermionic Choi state corresponding to $U$ is then
\begin{equation}
    \label{eq:positive_concurrence}
    \boxed{C_+(U,p) = \max\left(0, \; \frac{1}{2}\eta \left(1-p\right)\left(2-p\right) + \frac{p^2}{8} - \frac{p}{2} \right).}
\end{equation}
This is consistent with the first graph in Figure~\ref{fig:swap_graph}. There are a few things to notice about $\Delta^+$. Firstly, we can find the first derivative in $p$ to be
\begin{equation}
    \partial_p \Delta^+ = p \left( \frac{1}{4} + \eta \right) - \frac{3}{2} \eta - \frac{1}{2}.
\end{equation}
We can see that this derivative is a linear polynomial and is always negative for $p \in [0,1]$ and thus in that range, $\Delta^+$ is monotonically decreasing and has a single root. Further, we can also trivially see that $\eta=1$ will maximize $\Delta^+$. Therefore, the value of $p$ at which the positive concurrence becomes zero is maximized for $\eta=1$. With this, we can set Eq.~\eqref{eq:positive_delta} to zero and trivially find its root to be
\begin{equation}
    \boxed{p^+_\text{max} = \frac{8 - 2 \sqrt{6}}{5} \approx 0.6202.}
\end{equation}

\newpage

All of the four odd pair blocks have the values $(4p - 3p^2)/16$ and $p^2/16$ for $w_1$ and $w_2$. Thus, $\tau_-$ only has two unique singular values, $\sigma_+ \geq \sigma_-$. Therefore,
\begin{equation}
    \Delta^- = \sigma_+ - (3\sigma_+ + 4\sigma_-) = -2\sigma_+ - 4\sigma_- \leq0,
\end{equation}
and thus
\begin{equation}
    \label{eq:negative_concurrence}
    \boxed{C_-(U,p) = 0.}
\end{equation}

Basic calculation shows that SWAP has $\eta=1$, and so maximizes the noise it tolerates before its combined channel becomes convex Gaussian. This is also true for the CZ gate and any parity preserving gate where its block determinants are maximally far apart.

\section{Derivation of the W Operator}
\label{app:w_matrix_derivation}

We now derive the matrix form for the operator $W$. Reminding ourselves of the definition of $\ket{\psi}$, we can expand $\Omega_{\mu \nu}$ in Eq.~\eqref{eq:omega_expanded} to be
\begin{equation}
    \begin{aligned}
        \Omega_{\mu \nu} &= - \bra{\psi} (K_\mu \mathcal{J} K_\nu^* \otimes \mathcal{J}) (V_\delta \otimes I)^* \ket{\Phi_G}^* \\
        &= - \bra{\psi} (K_\mu \mathcal{J} K_\nu^* \otimes \mathcal{J}) (V_\delta^* \otimes R^*) \ket{\Phi_4}^*.
    \end{aligned}
\end{equation}
Assuming we're in the computational basis and using the transfer identity for maximally entangled bipartite states, we get
\begin{equation}
    \begin{aligned}
        \Omega_{\mu \nu} &= - \bra{\psi} (K_\mu \mathcal{J} K_\nu^* \otimes \mathcal{J}) (V_\delta^\dagger R^\dagger \otimes I) \ket{\Phi_4} \\
        &= - \bra{\psi} (K_\mu \mathcal{J} K_\nu^* \otimes I) (V_\delta^\dagger R^\dagger \mathcal{J}^T \otimes I) \ket{\Phi_4} \\
        &= - \bra{\psi} (K_\mu \mathcal{J} K_\nu^* \otimes I) (-V_\delta^\dagger R^\dagger \mathcal{J} \otimes I) \ket{\Phi_4}.
    \end{aligned}
\end{equation}
It is simple to show that
\begin{equation}
    \ket{\psi} = (V_\delta \otimes R) \ket{\Phi_4}= (V_\delta R^T \otimes I) \ket{\Phi_4},
\end{equation}
and thus
\begin{equation}
    \ket{\Phi_4} = (V_\delta R^T \otimes I)^{-1} \ket{\psi} = ((R^T)^{-1} V_\delta^\dagger \otimes I) \ket{\psi}.
\end{equation}
Given the matrix form for $R$, one can show this is equivalent to
\begin{equation}
    \ket{\Phi_4} = (-R^\dagger V_\delta^\dagger \otimes I) \ket{\psi}.
\end{equation}
Therefore,
\begin{equation}
    W = V_\delta^\dagger R^\dagger \mathcal{J} R^\dagger V_\delta^\dagger.
\end{equation}

It is straightforward to show that all of these operators commute among themselves. Thus, we can rewrite $W$ as
\begin{equation}
    W = \mathcal{J} (R^\dagger)^2 (V_\delta^\dagger)^2= \mathcal{J} P \exp(-i \delta P / 2),
\end{equation}
Using Euler's formula, this can be written as
\begin{equation}
    \begin{aligned}
        W &= \mathcal{J} P \left( \cos(-\frac{\delta}{2}) + i \sin(-\frac{\delta}{2}) P \right) \\
        &= \mathcal{J} \left( i \sin\frac{\delta}{2} - \cos\frac{\delta}{2} P\right ). \\
    \end{aligned}
\end{equation}

\clearpage

\bibliography{main.bib}

\end{document}